\documentclass[aps,prb,twocolumn,superscriptaddress]{revtex4-1}

\usepackage{graphicx}
\usepackage{hyperref}
\usepackage{url}
\usepackage{amsmath}
\usepackage{verbatim}
\usepackage{color}
\usepackage{bm}
\usepackage{amssymb}

\def\beq {\begin{equation}}
\def\eeq {\end{equation}}

\def\bfr {\mathbf{r}}
\def\bfR {\mathbf{R}}

\def\bfq {\mathbf{q}}

\def\bfG {\mathbf{G}}

\date{\today}

\begin{document}
\title{Separation between long- and short-range part of the Coulomb interaction in low dimensional systems: implications for the macroscopic screening length and the collective charge excitations.}

\author{Pierluigi Cudazzo}
\affiliation{Dipartimento di Fisica, Universit\'a di Trento, via Sommarive 14, I-38123 Povo, Italy}
\affiliation{European Theoretical Spectroscopy Facility (ETSF)}

\begin{abstract}

Collective charge excitations directly probed in electron energy loss and inelastic X rays scattering spectroscopies play a key role in different fields of condensed matter physics. Being induced by the long-range part of the Coulomb interaction between particles, in standard bulk systems they appear as well-defined features in the spectra associated to the zero crossing of the real part of the macroscopic dielectric function. However, this simple criterion cannot be used to identify collective excitations in low dimensional systems where the macroscopic dielectric function is not a well defined concept. In this work, we discuss how this problem can be traced back to the definition of the long-range Coulomb interaction and we show how the appropriate separation between long- and short-range Coulomb interaction allows one to correctly express the low dimensional macroscopic dielectric function in terms of microscopic quantities accessible in first-principles calculations. This allows disentangling collective charge excitations in low dimensional materials in analogy with what one does in standard bulk systems. In addition, we show how important macroscopic quantities, such as the screening length scale, can be extracted from full first principle calculations. As an illustrative
example we perform a study of the screening effects and the collective excitations in prototypical 2D materials including metals (NbSe$_2$) as well as semiconductors (BN and MoS$_2$).

\end{abstract}


\maketitle

\section{Introduction}

The dynamical charge-density response function is a fundamental observable in condensed matter physics quantifying the longitudinal excitations of a system directly probed in electron energy loss (EEL) and inelastic X rays scattering (IXS) spectroscopies, the ability of a material to screen charge, as well as its electronic compressibility. Moreover it is a key quantity in transport phenomena, charge density wave (CDW) and superconducting phase transitions since it provides the renormalization of the electron-electron\cite{onida2002,strinati1988} and electron-phonon\cite{giustino2017} interactions. The behaviour of the dynamical charge-density response function is mainly set by the collective excitations of the system that dominate the EEL and the IXS spectra at small and intermediate momentum transfer, in contrast with the electron-hole continuum directly linked to the individual quasi-particle excitations that become more and more important as the momentum transfer increases.

In particular, collective charge excitations are tightly linked to interesting phenomena in fundamental physics, being plasmon or exciton softening\cite{kogar2017,pasquier2018,lian,varsano2017,barborini2022} the precursor of the CDW excitonic insulator and exciton Bose-Einstein condensate\cite{kohn,keldysh1965,jerome}. They are involved in the anti-screening effect associated to the presence of ghost excitons\cite{dolgov1981,ichimaru1982,takada2005,takada2016,koslelo2023} and can provide a pairing mechanism alternative to the electron-phonon coupling in unconventional superconductivity\cite{davis1976,ihm1981,choi2024}.

Beside that, charge collective excitations such as plasmons in confined geometries have important technological applications in heat generation\cite{gorcov2007,baffou2013,brongersma2015}, photoacoustic imaging\cite{yang2007,chen2011}, photocatalysis and photovoltaics\cite{lee2012,zhou2012,smith2015} thanks to their high tunability and their peculiarity of strongly enhancing light-matter interaction\cite{maier2007,fei2012,abajo2013}.

In this context the ability to disentangle the collective behaviour in the spectra of charge excitations is essential. In general, in an extended system, a collective mode appears as an isolated pole of the charge response function, in contrast with the continuum of quasi particle excitations usually associated to a branch-cut. This observation allows identifying the collective excitations in standard bulk systems as the zero crossing of the real part of the macroscopic dielectric function. However, this criterion is not suitable in low dimensional ($\ell D$) systems due to the difficulty in properly defining the macroscopic dielectric function\cite{sottile2005}. As a consequence, the collective excitations in the spectra of $\ell D$ systems are usually interpreted through phenomenological models\cite{pitarke2007,mazzei2022} that are often difficult to link to the microscopic charge response function, which is the basic quantity accessible through first principles calculations performed in the framework of Time Dependent Density Functional Theory (TDDFT) and Many Body Perturbation Theory (MBPT)\cite{onida2002}.

In this article, starting from the expression of the microscopic charge response function, which is unambiguously defined independently from the dimensionality, we provide a derivation of the basic spectroscopic quantities in $\ell D$ systems that allow disentangling the collective excitations in a systematic way without resorting to the introduction of phenomenological models. In addition, we show how important quantities, such as the screening length scale, can be extracted from full first principle calculations. Our results are general and independent from the approximations used to evaluate the microscopic charge response function. As an illustrative example we performed a study of the screening effects and the collective excitations in prototypical 2D materials including metals (NbSe$_2$) as well as semiconductors (BN and MoS$_2$).

\section{Macroscopic quantities in low dimensional systems}

Before analyzing the charge response function in low dimensional systems, it is useful to briefly review the definition of the basic spectroscopic quantities in standard 3D periodic crystals. 
By definition, a macroscopic position-dependent quantity is characterized by a length scale very large respect to the proper length scale of the crystal which is set by its lattice constant. This means that inside the unit cell a macroscopic quantity is essentially constant. In a standard bulk system, its value $f_M(\bfR)$ at a given lattice point $\bfR$ is defined as the average of the corresponding microscopic quantity $f(\bfr)$ taken over a unit cell\cite{wiser1963}. As a consequence, the Fourier component of a macroscopic quantity $f_M(\bfq)$ is given by $\lim_{\bfG\rightarrow 0}f_{\bfG}(\bfq)$\cite{wiser1963}. Here $f_{\bfG}(\bfq)$ is the $\bfq+\bfG$ Fourier component of the corresponding microscopic quantity, with $\bfG$ and $\bfq$ denoting a reciprocal lattice vector and a wave vector inside the first Brillouin zone (BZ), respectively. Similarly, for a non local function $f(\bfr,\bfr')$ we have that $f_M(\bfq)=\lim_{\bfG\rightarrow 0}f_{\bfG\bfG}(\bfq)$, where $f_{\bfG\bfG'}(\bfq)$ is the double Fourier transform of $f(\bfr,\bfr')$ that can be seen as a matrix function of $\bfq$ in the $\bfG$ space. This is a direct consequence of the translational invariance of $f(\bfr,\bfr')$\cite{wiser1963,adler1962}. 

Let $\phi(\bfr,\omega)$ be a frequency dependent external longitudinal field acting on a 3D crystal, according to the linear response theory, the Fourier component of the microscopic induced charge density is given by the following expression: 

\begin{equation}\label{eqcharge}
\rho^{ind}_{\bfG}(\bfq,\omega) = \chi_{\bfG\mathbf{0}}(\bfq,\omega)\phi(\bfq,\omega),
\end{equation}
where $\chi_{\bfG,\bfG'}(\bfq,\omega)$ is the microscopic charge response function. Here we take into account the fact that $\phi(\bfr,\omega)$ is a macroscopic quantity (i.e., its Fourier transform involves only the $\bfG=\mathbf{0}$ component). Taking the macroscopic limit in Eq. \ref{eqcharge}, the macroscopic induced charge can be written as $\rho^{ind}_M(\bfq,\omega)=\chi_M(\bfq,\omega)\phi(\bfq,\omega)$ with $\chi_M(\bfq,\omega)$ denoting the macroscopic response function. This quantity is linked to the inverse macroscopic dielectric function $\epsilon^{-1}_M(\bfq,\omega)=1+\frac{4\pi}{|\bfq|^2}\chi_M(\bfq,\omega)$\cite{wiser1963,adler1962} and gives direct access to the EEL function and the cross section of the IXS scattering.

In analogy with the homogeneous electron gas, $\chi_{M}$ satisfies a scalar Dyson like equation involving only the long-range part of the Coulomb potential. Indeed, cutting the Coulomb potential in a long-range part $v_0(\bfq)=\frac{4\pi}{|\bfq|^2}$ and a short-range part $\bar{v}_{\bfG}(\bfq)$\cite{onida2002,strinati1988}:

\begin{equation}\label{srC3d}
\bar{v}_{\bfG}(\bfq) =
\left\{
\begin{array}{rl}
0 & \mbox{for } \bfG = \mathbf{0} \\
\frac{4\pi}{|\bfq+\bfG|^2}  & \mbox{for } \bfG \neq \mathbf{0}
\end{array}
\right.
\end{equation}  
the Dyson equation for the microscopic charge response function can be split in the following system of equations:

\begin{eqnarray}
\bar{\chi}_{\bfG\bfG'}(\bfq,\omega) &=& \pi_{\bfG\bfG'}(\bfq,\omega) \nonumber \\
&+& \sum_{\bar{\bfG}}\pi_{\bfG\bar{\bfG}}(\bfq,\omega)\bar{v}_{\bar{\bfG}}(\bfq)\bar{\chi}_{\bar{\bfG}\bfG'}(\bfq,\omega) \\
\chi_{\bfG\bfG'}(\bfq,\omega) &=& \bar{\chi}_{\bfG\bfG'}(\bfq,\omega) \nonumber \\
&+& \bar{\chi}_{\bfG\mathbf{0}}(\bfq,\omega)v_0(\bfq)\chi_{\mathbf{0}\bfG'}(\bfq,\omega),\label{eqchi3da}
\end{eqnarray}
where $\pi$ is the proper part of $\chi$ and $\bar{\chi}$ is the modified microscopic charge response function\cite{onida2002,strinati1988}. The macroscopic limit of Eq. \ref{eqchi3da} provides a scalar Dyson like equation for $\chi_M(\bfq,\omega)$ in terms of $\bar{\chi}_M(\bfq,\omega)=\bar{\chi}_{\mathbf{0}\mathbf{0}}(\bfq,\omega)$ and thus the spectrum of the longitudinal charge excitations:

\begin{equation}\label{eqchi3db}
\Im\chi_M(\bfq,\omega)=\frac{\Im\bar{\chi}_M(\bfq,\omega)}{[\Re\epsilon_M(\bfq,\omega)]^2+[\Im\epsilon_M(\bfq,\omega)]^2},
\end{equation}
where the denominator has been expressed in terms of the macroscopic dielectric function: $\epsilon_M(\bfq,\omega)=1-v_0(\bfq)\bar{\chi}_M(\bfq,\omega)$\cite{onida2002,strinati1988}. As can be inferred from Eq. \ref{eqchi3db}, the spectrum of the longitudinal charge excitations involves the electron-hole continuum with the effect of crystal local fields induced by the short range charge fluctuations which is described by $\Im\bar{\chi}_M$, as well as the isolated poles of $\chi_M$ that define the collective longitudinal charge excitations induced by the long range part of the Coulomb potential. They constitute the dominant excitations of the system and are defined by the zeros of the denominator of Eq. \ref{eqchi3db} set by the well known conditions: $\Re\epsilon_M(\bfq,\omega)=0$ and $\Im\epsilon_M(\bfq,\omega)\approx 0$.  


The situation is quite different in low dimensional ($\ell D$) systems with $\ell=1,2$, where a macroscopic average can be defined in this way\cite{mazzei2022,dahland1977,hogan2009}:

\begin{equation}\label{eqmacro}
f_M(\bfR)=\lim_{\Omega_\perp\rightarrow\infty}\frac{1}{\Omega_{\parallel}}\int_{\Omega_\parallel}d\bfr_\parallel\int_{\Omega_\perp}d\bfr_{\perp}f(\bfr_\parallel+\bfR,\bfr_\perp).
\end{equation}
Here $\bfr_\parallel$ and $\bfr_\perp$ denote the components of $\bfr=(\bfr_\parallel,\bfr_\perp)$ parallel and normal to the periodic direction, respectively and $\Omega_\parallel$ is the volume of the $\ell D$ unit cell so that it has the dimension of a surface for $\ell=2$ and a length for $\ell=1$. $\Omega_\perp$, on the other hand, is the volume associated to the non periodic directions. The fact that the electronic wave functions are confined along the non periodic directions ensures that the integral in Eq. \ref{eqmacro} well behaves for  $\Omega_\perp\rightarrow\infty$. The corresponding Fourier component is
\begin{equation} 
f_M(\bfq)=\lim_{\Omega_\perp\rightarrow\infty}\int_{\Omega_\perp}d\bfr_\perp f_{\bfG_\parallel=\mathbf{0}}(\bfq,\bfr_\perp)
\end{equation}
where $f_{\bfG_\parallel}(\bfq,\bfr_\perp)$ is the $\ell D$ partial Fourier transform of the corresponding microscopic quantity $f(\bfr)$.

Let $\phi(\bfr_\parallel,\omega)$ be a frequency dependent external potential that by definition is independent from $\bfr_\perp$, the microscopic induced charge in linear response approximation is given by the following expression\cite{mazzei2022,dahland1977,hogan2009}:

\begin{equation}
\rho^{ind}(\bfr_\parallel,\bfr_\perp;\omega)=\int d\bfr'_\parallel\int d\bfr'_\perp\chi(\bfr,\bfr'_\parallel,\bfr'_\perp;\omega)\phi(\bfr'_\parallel;\omega),
\end{equation}
so that the Fourier component of the corresponding macroscopic quantity becomes:

\begin{equation}
\rho^{ind}_M(\bfq,\omega)=\chi^{\ell D}_M(\bfq,\omega)\phi(\bfq,\omega)
\end{equation}
with $\chi^{\ell D}_M(\bfq,\omega)$ denoting the Fourier transform of the $\ell D$ macroscopic response function which is expressed in terms of the head of the $\ell D$ partial Fourier transform of $\chi(\bfr,\bfr';\omega)$:

\begin{equation}\label{eqchild1}
\chi^{lD}_M(\bfq,\omega)=\lim_{\Omega_\perp\rightarrow\infty}\int_{\Omega_\perp}d\bfr_\perp\int_{\Omega_\perp}d\bfr'_\perp\chi_{\mathbf{0}\mathbf{0}}(\bfq,\bfr_\perp,\bfr'_\perp;\omega).
\end{equation}
The integration of Eq. \ref{eqchild1} along the non periodic directions provides the explicit expression of $\chi^{\ell D}_M(\bfq,\omega)$ in terms of the head of the full Fourier transform of the microscopic response function: 
%

\begin{equation}\label{eqchildmacro} 
\chi^{lD}_M(\bfq,\omega)=\lim_{\Omega_\perp\rightarrow\infty}\Omega_\perp\chi_{\mathbf{0}\mathbf{0}}(\bfq,\omega).
\end{equation}
This is the basic spectroscopic quantity evaluated in TDDFT and MBPT based codes\cite{sangalli2019} and it is directly linked to the EEL function of a $\ell D$ system [$L^{\ell D}(\bfq,\omega)$], being $L^{\ell D}(\bfq,\omega)\propto-\frac{4\pi}{|\bfq|^2}\Im\chi^{lD}_M(\bfq,\omega)$.

In principle, in analogy with Eq. \ref{eqchildmacro}, we can introduce an equivalent expression for $\bar{\chi}^{\ell D}_M(\bfq,\omega)$. However, in contrast with the 3D case, $\chi^{\ell D}_M(\bfq,\omega)$ and $\bar{\chi}^{\ell D}_M(\bfq,\omega)$ are no more related to each other through a Dyson like equation since $\lim_{\Omega_\perp\rightarrow\infty}\bar{\chi}_{\mathbf{0}\mathbf{0}}(\bfq,\omega)=\lim_{\Omega_\perp\rightarrow\infty}\chi_{\mathbf{0}\mathbf{0}}(\bfq,\omega)$. This is a direct consequence of the fact that in $\ell D$ the non interacting response function that we use to build $\chi$ and $\bar{\chi}$ is proportional to $1/\Omega_\perp$\cite{sottile2005}. The lack of an equation equivalent to Eq. \ref{eqchi3db} that allows to disentangle in a simple way the electron-hole continuum makes in some cases extremely difficult the interpretation of the spectra in $\ell D$ systems. The problem can be traced back to the definition of the short range Coulomb interaction in Eq. \ref{srC3d} which is not appropriate in $\ell D$ and results in an ill defined $\bar\chi$ function.

\section{Long-range Coulomb interaction in low dimensions and local response approximation}

In the case of a $\ell D$ system it is convenient to separate a reciprocal lattice vector in a component parallel ($\bfG_\parallel$) and normal ($\bfG_\perp$) to the periodic direction, so that $\bfG=(\bfG_\parallel,\bfG_\perp)$ and the Fourier transform of the Coulomb potential takes the following structure: $v_{\bfG}(\bfq)=\frac{4\pi}{|\bfq+\bfG_\parallel+\bfG_\perp|^2}$. Since for $\Omega_\perp\rightarrow\infty$ the normal components of $\bfG$ follow a continuum distribution, the short range Coulomb interaction in Eq. \ref{srC3d} is not well defined for a $\ell D$ system, being $\bar{v}_{\bfG}(\bfq)=v_{\bfG}(\bfq)$. On the contrary, its appropriate definition is provided by the following expression\cite{mazzei2023}:

\begin{equation}\label{srCld}
\bar{v}_{\bfG_\parallel}(\bfq,\bfG_\perp) =
\left\{
\begin{array}{rl}
0 & \mbox{for } \bfG_\parallel = \mathbf{0} \\
\frac{4\pi}{|\bfq+\bfG_\parallel+\bfG_\perp|^2}  & \mbox{for } \bfG_\parallel \neq \mathbf{0}
\end{array}
\right.
\end{equation}
with the long-range part defined as $v_0(\bfq,\bfG_\perp)=\frac{4\pi}{|\bfq+\bfG_\perp|^2}$. In analogy with the 3D case the Dyson equation for the microscopic charge response function can be split in the following system of equations:

\begin{widetext}  
\begin{eqnarray}
\bar{\chi}^{\bfG_\perp\bfG_\perp'}_{\bfG_\parallel\bfG_\parallel'}(\bfq,\omega) &=& \pi^{\bfG_\perp\bfG_\perp'}_{\bfG_\parallel\bfG_\parallel'}(\bfq,\omega) + \sum_{\bar{\bfG}_\parallel\bar{\bfG}_\perp}\pi^{\bfG_\perp\bar{\bfG}_\perp}_{\bfG_\parallel\bar{\bfG}_\parallel}(\bfq,\omega)\bar{v}_{\bar{\bfG}_\parallel}(\bfq,\bar{\bfG}_\perp)\bar{\chi}^{\bar{\bfG}_\perp\bfG_\perp'}_{\bar{\bfG}_\parallel\bfG_\parallel'}(\bfq,\omega)\label{eqchia} \\
\chi^{\bfG_\perp\bfG_\perp'}_{\bfG_\parallel\bfG_\parallel'}(\bfq,\omega) &=& \bar{\chi}^{\bfG_\perp\bfG_\perp'}_{\bfG_\parallel\bfG_\parallel'}(\bfq,\omega)+\sum_{\bar{\bfG}_\perp}\bar{\chi}^{\bfG_\perp\bar{\bfG}_\perp}_{\bfG_\parallel\mathbf{0}}(\bfq,\omega)v_0(\bfq,\bar{\bfG}_\perp)\chi^{\bar{\bfG}_\perp\bfG_\perp'}_{\mathbf{0}\bfG_\parallel'}(\bfq,\omega),\label{eqchi3d}
\end{eqnarray}
where we introduced the notation $\chi^{\bfG_\perp\bfG_\perp'}_{\bfG_\parallel\bfG_\parallel'}(\bfq,\omega)=\chi_{\bfG\bfG'}(\bfq,\omega)$ (similarly for $\bar{\chi}$ and $\pi$) in order to separate the Fourier components with reciprocal wave vector parallel and normal to the periodic directions. From Eq. \ref{eqchi3d}, using the new definition of the modified microscopic response function $\bar{\chi}$ in Eq. \ref{eqchia}, we can write the expression for the head of the $\ell D$ partial Fourier transform of the charge response function:

\begin{equation}\label{eqchi3}
\chi_{\mathbf{0}\mathbf{0}}(\bfq,\bfr_\perp,\bfr'_\perp;\omega)= \bar{\chi}_{\mathbf{0}\mathbf{0}}(\bfq,\bfr_\perp,\bfr'_\perp;\omega)+\lim_{\Omega_\perp\rightarrow\infty}\int_{\Omega_\perp}d\bar{\bfr}_\perp\int_{\Omega_\perp}d\bar{\bar{\bfr}}_\perp\bar{\chi}_{\mathbf{0}\mathbf{0}}(\bfq,\bfr_\perp,\bar{\bfr}_\perp;\omega)v_0^{\ell D}(\bfq,|\bar{\bfr}_\perp-\bar{\bar{\bfr}}_\perp|)\chi_{\mathbf{0}\mathbf{0}}(\bfq,\bar{\bar{\bfr}}_\perp,\bfr'_\perp;\omega)
\end{equation}  
\end{widetext}
with $v_0^{\ell D}(\bfq,|\bfr_\perp-\bfr'_\perp|)$ denoting the $\ell D$ partial Fourier transform of the long range part of the Coulomb potential, so that:

\begin{equation}\label{srCld}
v^{\ell D}_0(\bfq,|\bfr_\perp-\bfr'_\perp|) =
\left\{
\begin{array}{rl}
\frac{2\pi}{|\bfq|}e^{-|\bfq||\bfr_\perp-\bfr'_\perp|} & \mbox{for } \ell = 2 \\
2K_0(|\bfq||\bfr_\perp-\bfr'_\perp|)  & \mbox{for } \ell = 1.
\end{array}
\right.
\end{equation}
where $K_0$ is the modified second kind Bessel function\cite{math1964}.

At this point it is important to note that, due to the localization of the electronic wave functions along the non-periodic directions, the polarizable charge is bound on the $\ell D$ system. This means that both $\chi$ and $\bar{\chi}$ are localized along the non-periodic directions as well. Therefore, let $\lambda$ be the thickness of the system, $\chi_{\mathbf{0}\mathbf{0}}(\bfq,\bfr_\perp,\bfr'_\perp;\omega)\approx 0$ for $|\bfr_\perp|>>\lambda$ and $|\bfr'_\perp|>>\lambda$, and similarly for the $\ell D$ partial Fourier transform of $\bar{\chi}$. This observation suggests the following ansatz for $\chi$ and $\bar{\chi}$:

\begin{eqnarray}
\chi_{\mathbf{0}\mathbf{0}}(\bfq,\bfr_\perp,\bfr'_\perp;\omega) &=& f(\bfq,\omega)|\phi(\bfr_\perp)|^2|\phi(\bfr'_\perp)|^2\label{lra1} \\
\bar{\chi}_{\mathbf{0}\mathbf{0}}(\bfq,\bfr_\perp,\bfr'_\perp;\omega) &=& \bar{f}(\bfq,\omega)||\phi(\bfr_\perp)|^2\phi(\bfr'_\perp)|^2\label{lra2},
\end{eqnarray}
which is based on the assumption that the electronic wave functions can be factorized in a component parallel and normal to the periodic direction of the system  and that $\chi$ and $\bar{\chi}$ are not strongly sensitive on the details of their spatial distribution in the normal directions. This is justified as long as the length scale of the macroscopic induced charge is very large respect to the decay length scale of $\phi(\bfr_\perp)$ (i.e. $|\bfq|\lambda <<1$) that can be interpreted as the normal component of the electronic wave function. In the following we will refer to Eqs. \ref{lra1} and \ref{lra2} as local response approximation (LRA). In addition, since $\phi(\bfr_\perp)$ is normalized, Eq. \ref{eqchild1} implies that $f(\bfq,\omega)=\chi^{\ell D}_M(\bfq,\omega)$ and  $\bar{f}(\bfq,\omega)=\bar{\chi}^{\ell D}_M(\bfq,\omega)$. 

Inserting Eqs. \ref{lra1} and \ref{lra2} into Eq. \ref{eqchi3} and taking the integral over $\bfr_\perp$ and $\bfr'_\perp$, we obtain a scalar Dyson equation for $\chi^{\ell D}_M(\bfq,\omega)$ in terms of $\bar{\chi}^{\ell D}_M(\bfq,\omega)$ so that:

\begin{equation}\label{eqchildb}
\Im\chi^{\ell D}_M(\bfq,\omega)=\frac{\Im\bar{\chi}^{\ell D}_M(\bfq,\omega)}{[\Re\epsilon^{\ell D}_M(\bfq,\omega)]^2+[\Im\epsilon^{\ell D}_M(\bfq,\omega)]^2},
\end{equation}  
where we have introduced the $\ell D$ macroscopic dielectric function $\epsilon^{\ell D}_M(\bfq,\omega)=1-v^{\ell D}_{eff}(\bfq)\bar{\chi}^{\ell D}_M(\bfq,\omega)$, with $v^{\ell D}_{eff}(\bfq)$ denoting the matrix element of the $\ell D$ partial Fourier transform of the long range Coulomb potential:

\begin{equation}
v^{\ell D}_{eff}(\bfq)=\int_{\Omega_\perp}d\bfr_\perp\int_{\Omega_\perp}d\bfr'_\perp v^{\ell D}_0(\bfq,|\bfr_\perp-\bfr'_\perp|)F(\bfr_\perp,\bfr'_\perp),
\end{equation} 
being $F(\bfr_\perp,\bfr'_\perp)=|\phi(\bfr_\perp)|^2|\phi(\bfr'_\perp)|^2$.

For example, taking a Gaussian distribution for the normal component of the electronic wave function $\phi(\bfr_\perp)=(\frac{2}{\pi\lambda})^{\frac{3-\ell}{4}}e^{-|\bfr_\perp|^2/\lambda^2}$, we have:

\begin{equation}\label{veffld}
v^{\ell D}_{eff}(\bfq) =
\left\{
\begin{array}{rl}
\frac{2\pi}{|\bfq|}[1-\text{erf}(\lambda|\bfq|)]e^{\lambda|\bfq|} & \mbox{for } \ell = 2 \\
-e^{\lambda^2|\bfq|^2}\text{Ei}(-\lambda^2|\bfq|^2)  & \mbox{for } \ell = 1,
\end{array}
\right.
\end{equation}    
with erf and Ei denoting the error function and the exponential integral function, respectively\cite{math1964}. From their asymptotic behaviour for $|\bfq|\lambda << 1$, we recover the expression of the 2D and 1D Coulomb potentials which are independent from the functional form of $\phi(\bfr_\perp)$: $v^{2D}_{eff}(\bfq)=\frac{2\pi}{|\bfq|}$ and $v^{1D}_{eff}(\bfq)=-2\ln(\lambda|\bfq|)$. 
Moreover, since at small wave vectors $\bar{\chi}^{\ell D}_M$ is proportional to $|\bfq|^2$, we have that $\lim_{\bfq\rightarrow 0}\chi^{\ell D}_M(\bfq,\omega)=\lim_{\bfq\rightarrow 0}\bar{\chi}^{\ell D}_M(\bfq,\omega)$ which is consistent with the lack of LO-TO splitting in $\ell D$ systems\cite{koskelo2017,villanueva2024}.

Summarizing, the separation between long- and short-range Coulomb interaction in Eq. \ref{srCld} provides the useful definition of the modified charge response function in $\ell D$ systems that allows writing a scalar Dyson equation for $\chi^{\ell D}_M$ in terms of $\bar{\chi}^{\ell D}_M$. In this way, starting from Eq. \ref{eqchildb}, in analogy with the $3D$ case, we can define the $\ell D$ collective excitations as the zeros of the real part of the macroscopic $\ell D$ dielectric function. In particular in the case of a metal $\lim_{\bfq\rightarrow 0}\bar{\chi}^{\ell D}_M(\bfq,\omega)\propto\frac{|\bfq|^2}{\omega^2}$ in agreement with the random phase approximation (RPA), which is formally exact in the optical limit\cite{vignale2005}. Insertion of this expression in the equation $\Re\epsilon^{\ell D}_M(\bfq,\omega)=0$, leads to the well known dispersion relation for the intra-band plasmon in the optical limit: $\omega^{2D}_P(\bfq)\propto\sqrt{|\bfq|}$\cite{vignale2005,cudazzo2013,andersen2013,jornada2020} and $\omega^{1D}_P(\bfq)\propto|\bfq||\ln(\lambda|\bfq|)|$\cite{vignale2005}. Similarly, for bound states (excitons) in semiconductors, the $\ell D$ nature of the long-range part of the Coulomb interaction is responsible for the following exciton dispersion relations: $\omega^{2D}_E(\bfq)=\omega_0+(\boldsymbol{\mu}\cdot\hat{\bfq})^2|\bfq|$\cite{cudazzo2016,qiu2015} and $\omega^{1D}_E(\bfq)=\omega_0+(\boldsymbol{\mu}\cdot\hat{\bfq})^2|\bfq|^2\ln(\lambda|\bfq|)$\cite{qiu2021}, $(\boldsymbol{\mu}\cdot\hat{\bfq})$ being the projection of the exciton dipole matrix element along $\bfq$.

From the expression of the long-range part of the Coulomb potential in Eq. \ref{srCld}, we clearly see that, in 2D systems, as long as $|\bfq|\lambda <<1$, we can safely neglect the exponential factor taking $|\bfr_\perp-\bfr'_\perp|=0$ and $v^{2D}_0(\bfq,|\bfr_\perp-\bfr'_\perp|)=\frac{2\pi}{|\bfq|}$ into Eq. \ref{eqchi3}. This means that, depending on the thickness of the material, we can always define a regime in which the LRA is formally exact. This is not the case in 1D systems due to the behaviour of the $K_0$ function that has a divergence at zero. As a consequence, in 1D, finite thickness effects must be always taken into account and the LRA is always an approximation. From a mathematical point of view, this means that the 1D Coulomb potential must be regularized. Therefore, while in 2D the macroscopic dielectric function is in principle a well defined quantity able to correctly describe screening effects on a length scale large respect to the thickness of the material, in 1D systems the correct description of the screening always requires a full microscopic treatment.

\section{Local response approximation in 2D}

\begin{figure}[t]
\includegraphics[width=1.0\columnwidth]{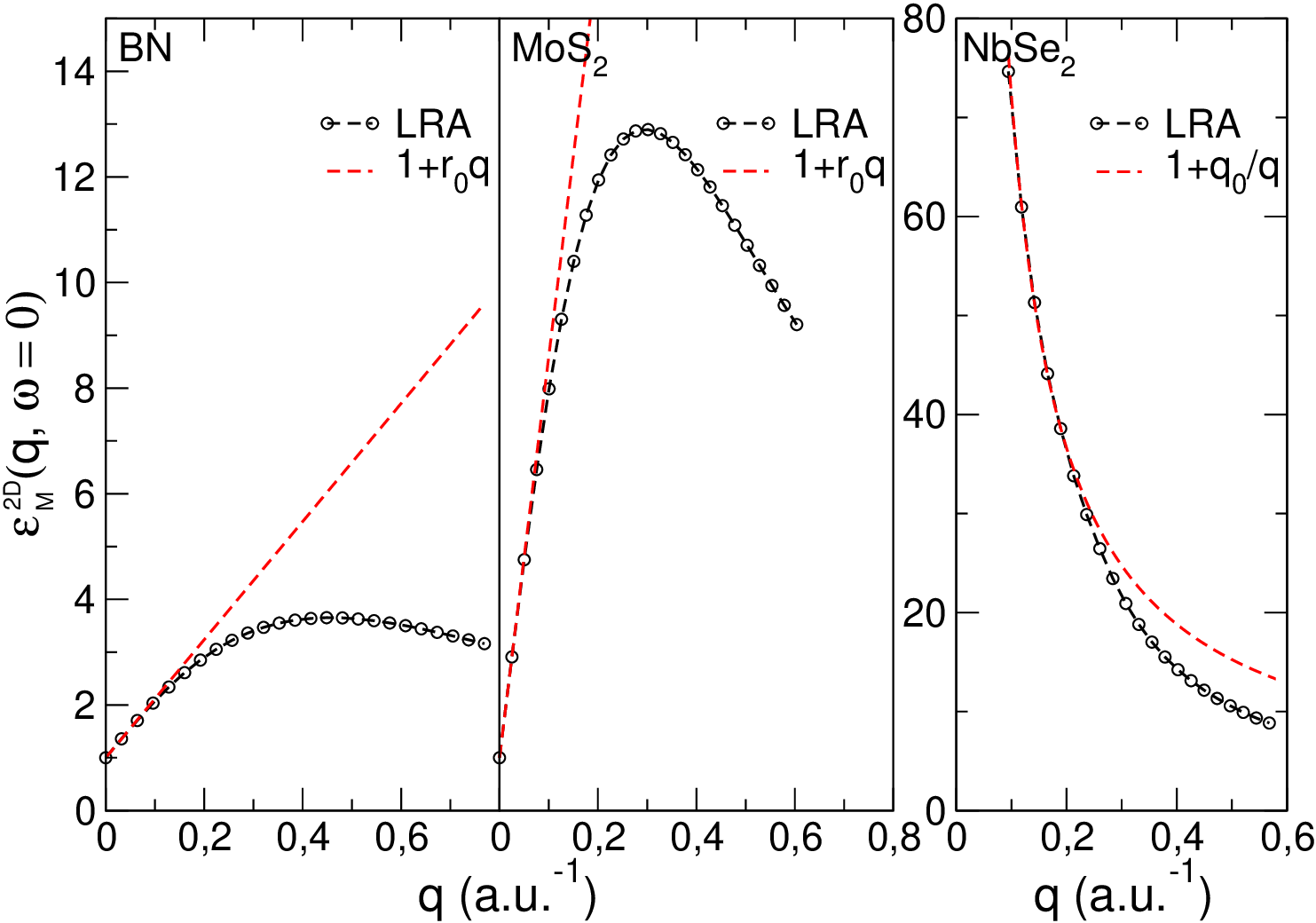}
\caption{Static 2D macroscopic dielectric constant for three prototypical 2D systems involving semiconductors (BN and MoS$_2$) and metals (NbSe$_2$).} 
\label{fig1}
\end{figure}

The analysis carried out so far has shown how the LRA is formally exact in 2D as long as $|\bfq|\lambda << 1$. This provides the definition of the 2D macroscopic dielectric function as:

\begin{equation}\label{eqeps2d}
\epsilon^{2D}_M(\bfq,\omega)=1-\frac{2\pi}{|\bfq|}\bar{\chi}^{2D}_M(\bfq,\omega),
\end{equation}
where $\bar{\chi}^{2D}_M$ is obtained from the solution of Eq. \ref{eqchia}. At shorter length-scale, such that $|\bfq|\lambda\gtrsim 1$, the LRA is no more exact even if finite thickness effects in Eq. \ref{eqeps2d} can be included using model effective interactions instead of the 2D Coulomb potential like in Eq. \ref{veffld} or more sophisticated approximations\cite{latini2015}. Under these conditions, in analogy with the 1D case, the exact description of the screening requires a full microscopic treatment. Therefore, in the following we will focus in the regime $|\bfq|\lambda << 1$, where Eqs. \ref{eqeps2d} and \ref{eqchildb} are nearly exact.

In Fig. \ref{fig1} we report the behaviour of the static dielectric function $\epsilon^{2D}_M(\bfq,\omega=0)$ evaluated in RPA as a function of the wave vector for BN, MoS$_2$ and NbSe$_2$. In the case of  semiconductors, it displays a linear dependence for small $\bfq$ and goes to 1 at $\bfq=0$ according with the well known behaviour predicted solving the Poisson's equation for a 2D macroscopically screened Coulomb potential\cite{cudazzo2011,cudazzo2010}: $\lim_{\bfq\rightarrow\infty}\epsilon^{2D}_M(\bfq,\omega=0)=1+r_0|\bfq|$. Here $r_0$ is the screening length that quantifies the ability of a 2D dielectric to screen charge in analogy with the dielectric constant in bulk systems. Direct comparison with Eq. \ref{eqeps2d} leads to an explicit expression in atomic units (a.u.) of the screening length in terms of the 2D modified response function that can be evaluated ab-initio: 

\begin{equation}
r_0=\lim_{\bfq\rightarrow 0}-\frac{2\pi}{|\bfq|^2}\bar{\chi}^{2D}_M(\bfq,\omega=0).
\end{equation}
In particular, for BN and Mo$S_2$ we find a screening length of 11.2 a.u. and 76.0 a.u., respectively. 

For a metallic system $\lim_{\bfq\rightarrow 0}\epsilon^{2D}_M(\bfq,\omega=0)=1+\frac{q_0}{|\bfq|}$, with $q_0$ denoting the 2D Thomas-Fermi screening wave vector\cite{ando1982}. Comparison with Eq. \ref{eqeps2d} (see Fig. \ref{fig1} right panel) allows expressing $q_0$ in a.u.$^{-1}$ in terms of $\bar{\chi}^{2D}_M$ as well:

\begin{figure}[t]
\includegraphics[width=0.8\columnwidth]{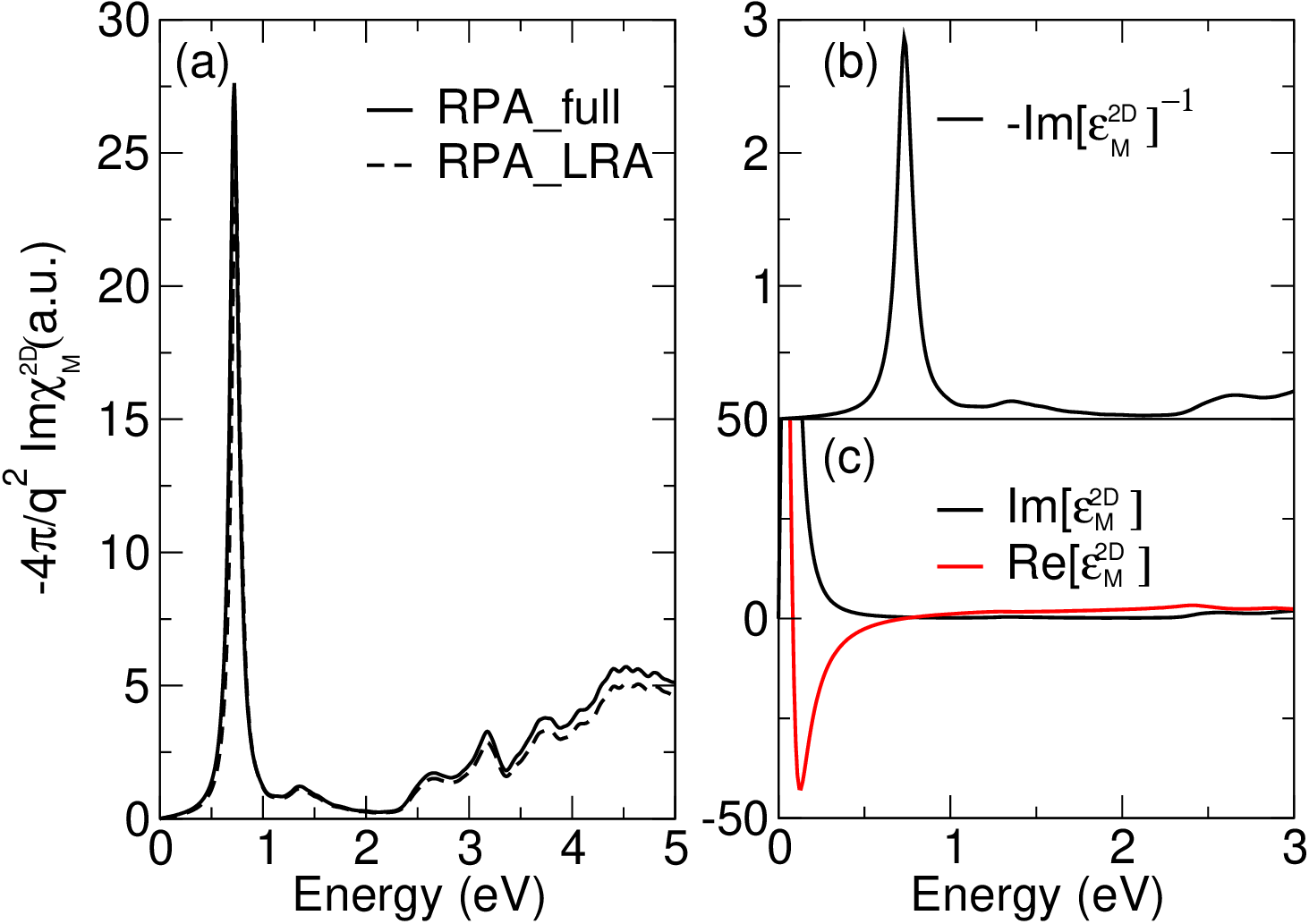}
\caption{NbSe$_2$ spectra at $q=0.02$ a.u.$^{-1}$ along the $\Gamma M$ direction evaluated in RPA: (a) comparison between the imaginary part of the 2D macroscopic response function evaluated through the full inversion of the microscopic Dyson equation (RPA full) and in LRA (RPA LRA); (b) imaginary part of the 2D inverse macroscopic dielectric function; (c) real and imaginary part of the 2D macroscopic dielectric function.}  
\label{fig2}
\end{figure}

\begin{equation} 
q_0=-\lim_{\bfq\rightarrow 0}2\pi\bar{\chi}^{2D}_M(\bfq,\omega=0),     
\end{equation}
giving for NbSe$_2$ the value $q_0=7.1$ a.u.$^{-1}$.

In Fig. \ref{fig2} (a) we compare the imaginary part of the RPA dynamical charge response function of NbSe$_2$ at $\bfq=0.02$ a.u.$^{-1}$ along the $\Gamma$M direction evaluated in LRA and through the solution of the full microscopic Dyson equation. The two curves are essentially indistinguishable except for a small discrepancy in the high frequency region of the spectrum related to finite thickness effects that, at this wave vector, are not completely negligible. As a matter of fact,  rather than a single monolayer like graphene and BN, TMDs are three layers systems involving a layer of transition metal atoms encapsulated between two layers of chalcogen atoms. As a consequence, the pure 2D behaviour can be observed only at very small wave vectors. The spectrum of NbSe$_2$ is dominated by a sharp peak at 0.72 eV that, due to the metallic nature of this material, can be intuitively ascribed to a plasmon. Nevertheless, based on the knowledge of the spectrum alone, it is not obvious to establish the nature of this excitation. Only from the analysis of the 2D macroscopic dielectric function in Fig. \ref{fig2} (b) and (c) we can conclude that it is a 2D intraband plasmon being $\Re\epsilon^{2D}_M(\bfq,\omega)=0$ at that frequency. In addition, through the study of the 2D macroscopic dielectric function  we can gain further insight on the nature of the collective excitations since it allows to disentangle the effect of the thickness, interband transitions as well as crystal local fields.

\begin{figure}[t]
\includegraphics[width=0.8\columnwidth]{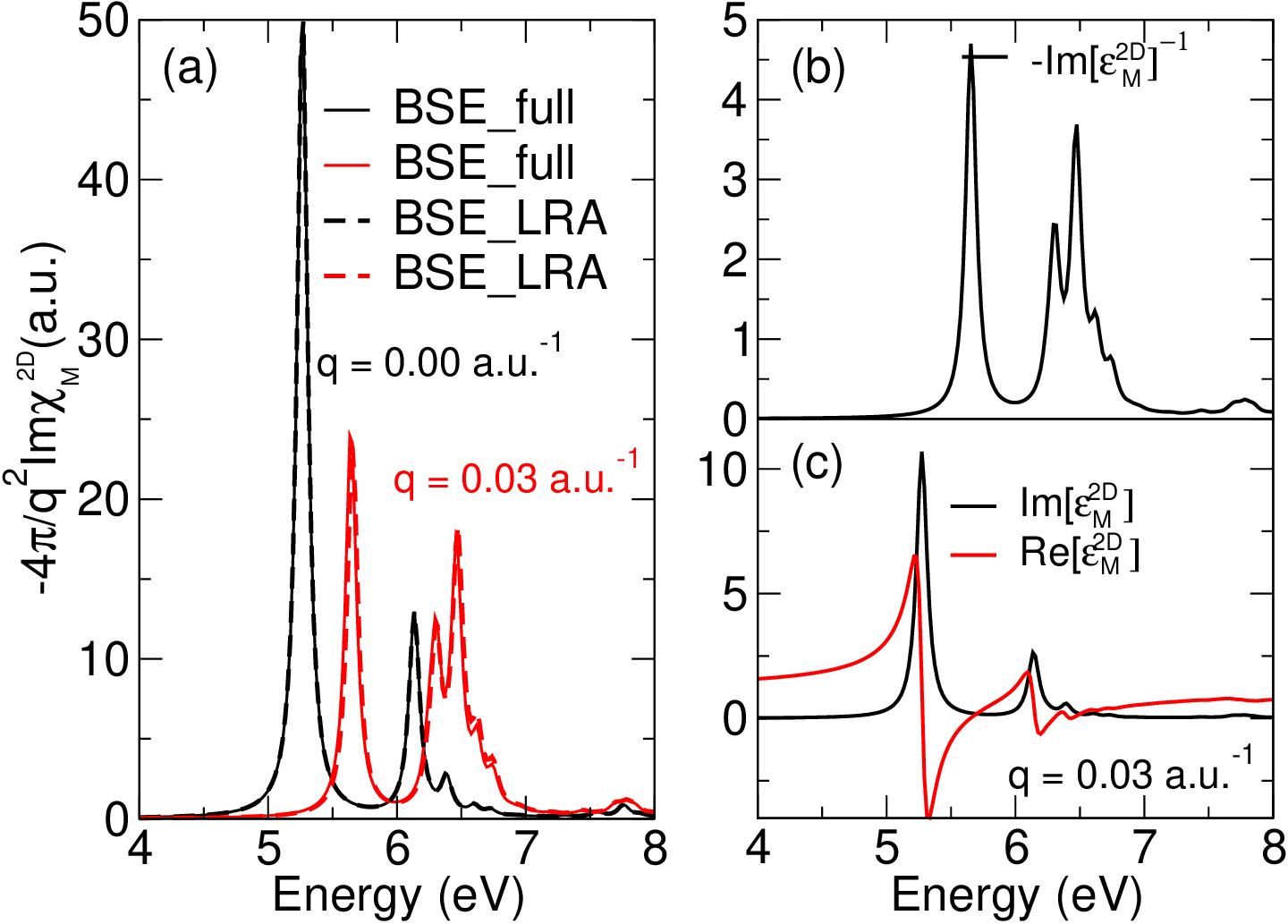}
\caption{BN spectra at different $q$ along the $\Gamma M$ direction evaluated in BSE: (a) comparison between the imaginary part of the 2D macroscopic response function evaluated through the full solution of the BSE (BSE full) and in LRA (BSE LRA); (b) imaginary part of the 2D inverse macroscopic dielectric function; (c) real and imaginary part of the 2D macroscopic dielectric function.} 
\label{fig3}
\end{figure}

So far we focused on the RPA where $\pi$ in Eq. \ref{eqchia} is treated at the independent particle level. However the LRA is general and can be employed to investigate excitonic effects in semiconductors as well. This requires the solution of the Bethe-Salpeter equation (BSE) for $\chi$ or in other words the evaluation of $\pi$ in the static ladder approximation\cite{onida2002,strinati1988}. As an illustrative example, we compare in Fig. \ref{fig3} (a) the 2D dynamical charge density response function of BN at different $\bfq$ evaluated in LRA and through the solution of the full BSE. 

The agreement between the two spectra is perfect as expected at small wave vectors. Interestingly, the behaviour of the 2D macroscopic dielectric function in Fig, \ref{fig3} (b) and (c) clearly shows how the energy of the lowest bright exciton correspond to a zero of $\Re\epsilon^{2D}_M(\bfq,\omega)$ highlighting the pure collective nature of this excitation in analogy with the intraband plasmon of NbSe$_2$. Nevertheless, it is important to note that the nature of the two collective excitations is completely different. Indeed, while in the case of the plasmon the coherent superposition of the electron-hole pairs responsible of the collective behaviour is induced by the long-range part of the Coulomb potential acting as an exchange electron-hole interaction, in the case of a longitudinal bound exciton the electron-hole superposition is mainly induced by the direct electron-hole interaction described by the statically screened Coulomb potential. This means that the nature of the first peak at 5.66 eV in $\Im[\epsilon^{2D}_M(\bfq,\omega)]^{-1}$ (Fig. \ref{fig3} (b)) and the first peak at 5.28 eV in $\Im\epsilon^{2D}_M(\bfq,\omega)$ (Fig. \ref{fig3} (c)) is essentially the same, so that the effect of the long-range part of the Coulomb interaction is to induce a blue-shift of the exciton energy without modifying remarkably the excitonic wave function, at last at small $\bfq$. This pure collective behaviour of the bound exciton is a peculiarity of materials characterized by strong excitonic effects that lead to a mixing of independent electron-hole pairs transitions on a wide region of the first BZ. This gives rise to sharp and isolated peaks in the $\Im\epsilon^{2D}_M(\bfq,\omega)$ well below the electronic band gap resulting in wide and fast oscillations of $\Re\epsilon^{2D}_M(\bfq,\omega)$ that eventually crosses the zero.                   

\section{Conclusions}

In conclusion, starting from the Dyson equation for the microscopic charge response function, we have shown that, in $\ell D$ systems, through an appropriate separation between long and short range Coulomb interaction it is possible to write a scalar Dyson equation for the macroscopic response function in terms of the modified response function associated to crystal local fields in analogy with what one does in standard bulk systems. In 2D, this equation is formally exact on a length scale large respect to the thickness of the system and allows one to identify charge collective excitations in a simple and intuitive way as zero crossing of the real part of the 2D macroscopic dielectric function. Finally we provided an expression of the screening length, that quantifies the ability of a 2D material to screen charge on a macroscopic scale, in terms of the modified charge response function that can be evaluated through first principle calculations. Our result is completely general being independent from the approximation used to evaluate the charge response function.




\begin{thebibliography}{100}

\bibitem{onida2002} G. Onida, L. Reining, and A. Rubio, Rev. Mod. Phys. {\bf 74}, 601 (2002).

\bibitem{strinati1988} G. Strinati, Riv. Nuovo Cimento {\bf 11}, 1 (1988).

\bibitem{giustino2017} F. Giustino, Rev. Mod. Phys. {\bf 89}, 015003 (2017).

\bibitem{kogar2017} A. Kogar, M. S. Rak, S. Vig, A. A. Husain, F. Flicker, Y. I. Joe, L. Venema, G. J. MacDougall, T. C. Chiang, E. Fradkin, J. van Wezel, and P. Abbamonte, Science {\bf 358}, 1314 (2017).

\bibitem{pasquier2018} D. Pasquier and O. V. Yazyev, Phys. Rev. B {\bf 98}, 235106 (2018).

\bibitem{lian} C. Lian, Z. A. Ali, and B. M. Wong, Phys. Rev. B {\bf 100}, 205423
(2019).

\bibitem{varsano2017} D. Varsano, S. Sorella, D. Sangalli, M. Barborini, S. Corni, E. Molinari, and M. Rontani, Nat. Commun. {\bf 8}, 1461 (2017).

\bibitem{barborini2022} M. Barborini, M. Calandra, F. Mauri, L. Wirtz, and P. Cudazzo, Phys. Rev. B {\bf 105}, 075122 (2022).


\bibitem{kohn} W. Kohn, Phys. Rev. Lett. {\bf 19}, 439 (1967).

\bibitem{keldysh1965} L. V. Keldysh and Y. V. Kopaev, Sov. Phys. Solid State. {\bf 6}, 2219 (1965).

\bibitem{jerome} D. J\'erome, T. M. Rice, and W. Kohn, Phys. Rev. {\bf 158}, 462
(1967).

\bibitem{dolgov1981} O. V. Dolgov, D. A. Kirzhnits, and E. G. Maksimov, Rev. Mod. Phys. {\bf 53}, 81 (1981).

\bibitem{ichimaru1982} S. Ichimaru, Rev. Mod. Phys. {\bf 54}, 1017 (1982).

\bibitem{takada2005} Y. Takada, Journal of Superconductivity {\bf 18}, 785 (2005).

\bibitem{takada2016} Y. Takada, Physical Review B {\bf 94}, 245106 (2016).

\bibitem{koslelo2023} J. Koskelo, L. Reining, M. Gatti, arXiv:2301.00474v1

\bibitem{davis1976} D. Davis, H. Gutfreund, and W. A. Little, Phys. Rev. B {\bf 13}, 4766 (1976).

\bibitem{ihm1981} J. Ihm, M. L. Cohen, and S. F. Tuan, Phys. Rev B {\bf 23}, 3258 (1981).

\bibitem{choi2024} Y. W. Choi, J. Ihm, and M. L. Cohen, Phys. Rev. B {\bf 110}, 155127 (2024).

\bibitem{gorcov2007} A. O. Govorov, and H. H. Richardson, Nano Today {\bf 2}, 30 (2007).

\bibitem{baffou2013} G. Baffou, and R. Quidant, Laser Photon. Rev. {\bf 7}, 171 (2013).

\bibitem{brongersma2015} M. L. Brongersma, N. J. Halas, and P. Nordlander, Nat. Nanotech. {\bf 10}, 25 (2015).

\bibitem{yang2007} X. Yang, et al., Nano Lett. {\bf 7}, 3798 (2007).

\bibitem{chen2011} Y. Chen, et al., Nano Lett. {\bf 11}, 348 (2011). 

\bibitem{lee2012} J. Lee, S. Mubeen, X. Ji, G. D. Stucky, and M. Moskovits, Nano Lett. {\bf 12}, 5014 (2012).

\bibitem{zhou2012} X. Zhou, G. Liu, J. Yu, and W. Fan, J. Mater. Chem. {\bf 22}, 21337 (2012).

\bibitem{smith2015} J. G. Smith, J. A. Faucheaux, and P. K. Jain, Nano Today {\bf 10}, 67 (2015).

\bibitem{maier2007} S. A. Maier "Plasmonics: Fundamental and Applications" (2007) (New York: Springer).

\bibitem{fei2012} Z. Fei, et al., Nature {\bf 487}, 82 (2012).
F. H. L. Koppens, D. E. Chang, F. J. Garc\'ia de Abajo, Nano Lett. {\bf 11}, 3370 (2011).

\bibitem{abajo2013} F. J. Garc\'ia de Abajo, Science {\bf 339}, 917 (2013).

\bibitem{sottile2005} F. Sottile, F. Bruneval, A. G. Marinopoulos, L. K. Dash, S. Botti, V. Olevano, N. Vast, A. Rubio, and L. Reining, Int. J. Quantum Chem. {\bf 102}, 684 (2005).

\bibitem{pitarke2007} J. M. Pitarke, V. M .Silkin, E. V. Chulkov, and P. M. Echenique, Rep. Prog. Phys. {\bf 70}, 1-87 (2007).

\bibitem{mazzei2022} S. Mazzei, and C. Giorgetti, Phys. Rev. B {\bf 106}, 035431 (2022).

\bibitem{wiser1963} N. Wiser, Phys. Rev. {\bf 129}, 62 (1963).

\bibitem{adler1962} S. L. Adler, Phys. Rev. {\bf 126}, 413 (1962). 

\bibitem{dahland1977} D. A. Dahland, and  L. J. Sham, Phys. Rev. B {\bf 17}, 641 (1977).

\bibitem{hogan2009} C. Hogan, M. Palummo, and R. D. Sole: Comptes Rendus Physique, {\bf 10}, 560 (2009).

\bibitem{sangalli2019} D. Sangalli et al., J. Phys.: Condens. Matter {\bf 31} 325902 (2019).

\bibitem{mazzei2023} S. Mazzei, and C. Giorgetti, Phys. Rev. B {\bf 107}, 165412 (2023).

\bibitem{math1964} "Handbook of Mathematical Functions With Formulas, Graphs, and Mathematical Tables", edited by M. Abramowitz and I. A. Stegun, NBS Applied Mathematics Series 55 (National Bureau of Standards, Washington, 1964).

\bibitem{koskelo2017} J. Koskelo, G. Fugallo, M. Hakala, M. Gatti, F. Sottile, and P. Cudazzo, Phys. Rev. B {\bf 95}, 035125 (2017).

\bibitem{villanueva2024} J. Cervantes-Villanueva, F. Paleari, A. Garc\'ia-Crist\'obal, D. Sangalli, and A. Molina-S\'anchez, Phys. Rev. B {\bf 109}, 155133 (2024).

\bibitem{vignale2005} G. Giuliani, and G. Vignale, "Quantum Theory of the Electron Liquid (Cambridge: Cambridge University Press) (2005). 

\bibitem{cudazzo2013} P. Cudazzo, M. Gatti, and A. Rubio, New Journal of Physics {\bf 15}, 125005 (2013).

\bibitem{andersen2013} K. Andersen, and K. S. Thygesen, Phys. Rev. B {\bf 88}, 155128 (2013).

\bibitem{jornada2020} F. H. da Jornada, L. Xian, A. Rubio, and S. G. Louie, Nature Comm. {\bf 11}, 1013 (2020).

\bibitem{cudazzo2016} P. Cudazzo, L. Sponza, C. Giorgetti, L. Reining, F. Sottile, and M. Gatti, Phys. Rev. Lett. {\bf 116}, 066803 (2016).

\bibitem{qiu2015} D. Y. Qiu, T. Cao, and S. G. Louie, Phys. Rev. Lett. {\bf 115}, 176801 (2015).

\bibitem{qiu2021} D. Y. Qiu, G. Cohen, D. Novichkova, and S. Refaely-Abramson, Nano letters {\bf 21}, 7644 (2021).

\bibitem{latini2015} S. Latini, T. Olsen, and K. S. Thygesen, Phys. Rev. B {\bf 92}, 245123 (2015).

\bibitem{cudazzo2011} P. Cudazzo, I. V. Tokatly, and A. Rubio, Phys. Rev. B {\bf 84}, 085406 (2011).

\bibitem{cudazzo2010} P. Cudazzo, C. Attacalite, I. V. Tokatly, and A. Rubio, Phys. Rev. Lett. {\bf 104}, 226804 (2010).

\bibitem{ando1982} T. Ando, A. B. Fowler, and F. Stern, Rev. Mod. Phys. {\bf 54}, 437 (1982).

\end{thebibliography}
\end{document}